\documentclass[a4paper]{article}

\usepackage{crop}
\usepackage{graphicx}
\usepackage{caption}
\usepackage{amsmath}
\usepackage{array}
\usepackage{color}
\usepackage{xcolor}
\usepackage{amssymb}
\usepackage{flushend}
\usepackage{stfloats}
\usepackage{chngpage}
\usepackage{totcount}
\usepackage{fix-cm}

\usepackage{hyperref}
\usepackage{listings}
\usepackage{booktabs}

\usepackage{authblk}

\usepackage{xspace}
\newcommand{\chemrecon}{ChemRecon\xspace}
\newcommand{\te}[1]{\emph{#1}}
\newcommand{\tr}[1]{\emph{#1}}

\begin{document}

\title{
	ChemRecon: a Consolidated Meta-Database Platform for Biochemical Data Integration
}

\author[1,$\ast$]{Casper Asbjørn Eriksen}
\author[1]{Jakob Lykke Andersen}
\author[1]{Rolf Fagerberg}
\author[2,1]{Daniel Merkle}

\affil[1]{
	Department of Mathematics and Computer Science,
	University of Southern Denmark
}
\affil[2]{
	Faculty of Technology,
	Bielefeld University
}
\date{}




\maketitle

\section*{Abstract}
	\textbf{Summary:}
	In this paper, we present \chemrecon, a meta-database and Python interface for integrating and exploring biochemical data across multiple heterogenous resources by consolidating compounds, reactions, enzymes, molecular structures, and atom-to-atom maps from several major databases into a single, consistent ontology.
	\chemrecon enables unified querying, cross-database analysis, and the construction of graph-based representations of sets of related database entries by the traversal of inter-database  connections. This facilitates information extraction which is impossible within any single database, including deriving consensus information from conflicting sources, of which identifying the most probable molecular structure associated with a given compound is just one example.
	\\
	\textbf{Availability and Implementation:} The Python interface is available via \texttt{pip} from the Python Package Index (\href{https://pypi.org/project/chemrecon/}{pypi.org/project/chemrecon/}). \chemrecon is open-source and the source code is hosted at GitLab (\href{https://gitlab.com/casbjorn/chemrecon}{gitlab.com/casbjorn/chemrecon}).
	\\
	\textbf{Contact:} Casper Asbjørn Eriksen: casbjorn@imada.sdu.dk
	\\
	\textbf{Supplementary Information:} Documentation and additional information is available at \href{https://chemrecon.org}{chemrecon.org}.

\section{Introduction}
\label{sec:intro}

Researchers increasingly rely on data in biochemical databases as core sources of knowledge.
A diverse ecosystem of specialized databases has emerged, each with their own focus.
Combining data from multiple such sources can, in principle, provide a far more comprehensive fpiview of the state of biochemical knowledge than any single database alone.

In practice, however, several obstacles prevent scientists from making use of the full potential of the vast amount of available data.
First, interfacing with each database typically requires a custom implementation for web access or for parsing database downloads, as data formats vary widely.
This makes working with multiple data sources a time-consuming process and hinders data integration.
As an example, many databases provide cross-references to related entries in other resources, but making effective use of these links requires accessing several sources, which carries the aforementioned challenges.
Second, many databases encode ontological relationships, but these are not compatible across resources, limiting their usefulness in settings where several data sources are used.
Third, the landscape of biochemical databases is not only heterogenous in the sense that they are formatted and accessed in different ways; they also frequently disagree with each other.
Discrepancies include listing different tautomers and identifying a completely different entry in another database as equivalent.

These challenges are apparent in common bioinformatics workflows.
For example, genome-scale metabolic models from the BiGG database offer a view of metabolites and reactions within organisms, but does not provide chemical or structural details on the metabolites themselves.
Instead, entries in BiGG are connected to other sources where this information is available e.g.\ MetaNetX, ChEbI, but exploiting these connections require nontrivial efforts to reconcile and integrate the sources.

Here we present \chemrecon, a consolidated meta-database with a Python interface designed to simplify the integration and exploration of biochemical data from a range of sources.
\chemrecon is built from full-database downloads of compounds, reactions, enzymes, molecular structures, and atom-to-atom maps from the following source databases:
BiGG~\cite{bigg},
BRENDA~\cite{brenda},
ChEbI~\cite{chebi},
ECMDB~\cite{ecmdb},
M-CSA~\cite{mcsa},
MetaMDB~\cite{metamdb},
MetaNetX~\cite{metanetx},
and PubChem~\cite{pubchem}
(see Table~\ref{tab:datasource}).
Heterogenous data formats were standardized, and relationships within and between these databases were reconstructed in a consistent format.
The resulting meta-database is freely accessible online, and is complemented by a Python interface which allows for easy integration into existing workflows.
This enables unified querying of entries from all the source databases, and discovery and visualization of relationships between these entries.
In contrast to existing integration resources (e.g., MetaNetX~\cite{metanetx}) which provide fixed identifier mappings, \chemrecon preserves the original database entries and exposes their cross-references as an explicit, traversable graph.
This design enables systematic cross-database exploration, complex cross-resource querying, and the derivation of consensus information from conflicting annotations across databases.
In short, \chemrecon simplifies workflows by allowing researchers to focus on scientific analyses rather than database engineering and enables knowledge discovery through its ability to construct and visualize graphs of associated biochemical information.

This paper describes the design and functionality of \chemrecon, presents practical examples of the use of the Python interface, and discusses potential applications.

%
%
\section{Description}
\label{sec:description}
\chemrecon consists of two main components: a consolidated meta-database and a corresponding Python interface which enables easy programmatic access to the database.
In this section, we describe the design and construction of the meta-database, the methods enabled by \chemrecon, and the usage of the interface.

\subsection{Data Sources}
\label{sec:datasources}

The \chemrecon meta-database was created from full downloads of the source databases. The contents of these downloads were then parsed and converted into a uniform format.
An overview of the data collected from the source databases is provided in Table~\ref{tab:datasource}.
The parsing and conversion routines are extensible, allowing users to expand the database capabilities by writing their own parsing scripts in case they have access to proprietary sources of data.
References from the source databases to other databases, including KEGG~\cite{kegg} and MetaCyc~\cite{metacyc}, are also included (with no additional information), allowing workflows based on identifiers from this larger set of databases.

\begin{table}
	\centering
	\scriptsize
	\begin{tabular}{@{}lrrrrr@{}}
		\toprule
		Source
		         & \te{Compound} & \te{MolStructure} & \te{Reaction} & \te{AAM} & \te{Enzyme} \\
		\midrule
		BiGG     & 20428         & -                 & 33942         & -        & 5705        \\
		BRENDA   & -             & -                 & 61129         & -        & 8697        \\
		ChEBI    & 224485        & 330207            & -             & -        & -           \\
		ECMDB    & 3760          & 3760              & -             & -        & -           \\
		M-CSA    & -             & -                 & 1003          & 342      & 1003        \\
		MetaMDB  & 80815         & 4392              & 74520         & 1003     & -           \\
		MetaNetX & 2601834       & 2297518           & 143880        & -        & 48175       \\
		PubChem  & 9031498       & 5000000           & -             & -        & -           \\
		\hline
	\end{tabular}
	\caption{
		The source databases contributing to the ChemRecon meta-database, and the number of entries sourced from each.
	}
	\label{tab:datasource}
\end{table}

\subsection{Database Structure}
\label{sec:db}

Each entry present in the source databases are consolidated into the \chemrecon meta-database as an \emph{entry}.
The meta-database contains various \emph{entry types}, including \te{Compound}, \te{Reaction}, and \te{Enzyme}, \te{MolStructure}, and \te{AAM}.
Within each entry type, entries have different identifiers corresponding to the source of the entry, represented by the \emph{source id} and \emph{id type}, which together forms a reference to the source of the entry.
For example, the MetaNetX \te{Compound} entry for `glucose' has a source id of `\verb|MNXM7381|' and an id type of `\verb|MetaNetX Compound|'.

References between database entries are encoded in the database as \emph{relations}.
As with entries, relations have \emph{relation types}, each type representing a distinct category of relation between two entry types.
The following is a non-exhaustive list of relation types, representing the diversity of connections in the \chemrecon database:
\begin{itemize}
	\item \tr{CompoundReference}
	      ($\te{Compound} \leftrightarrow \te{Compound}$):
	      encodes a reference between compound entries in different source databases, representing equivalent or similar compounds, with exact semantics differing depending on the source of the reference.
	\item \tr{CompoundParticipatesInReaction} / \tr{ReactionHasParticipant} \\
	      ($\te{Compound} \rightarrow \te{Reaction}$):
	      connects the compounds which take part in a reaction and contains the stoichiometric coefficient.
	\item \tr{Standardize}
	      ($\te{MolStructure} \rightarrow \te{MolStructure}$):
	      represents a standardization operation performed on molecular structures. This can be used to reconcile structures with respect to features such as charge, tautomerism, and stereoisomerism, using methods described in \cite{structrecon}.
	\item \tr{HasInstance} / \tr{IsA}
	      (e.g., $\te{Compound} \rightarrow \te{Compound}$):
	      these relations are gathered from information in the source databases, where they are typically used when a compound, reaction, or enzyme entry is considered a `class'.
	      For instance, in ChEBI, `isocitrate' \emph{is a} `tricarboxylic acid'.
\end{itemize}

A complete list of types of entries and relations with further details can be found in the online documentation (\href{chemrecon.org}{chemrecon.org}).

\begin{figure}
	\centering
	\includegraphics[scale=0.3]{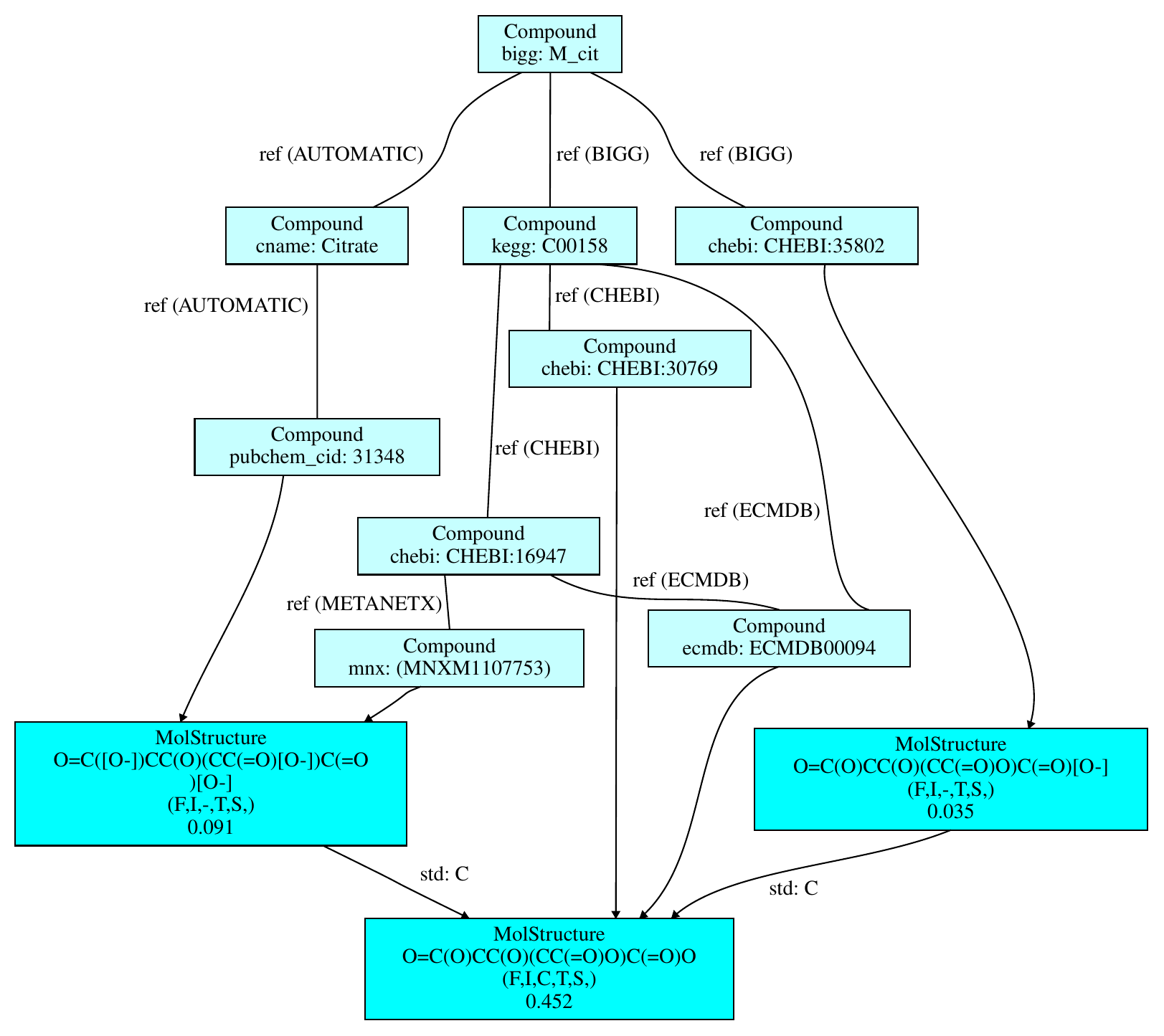}
	\caption{
		An excerpt of the \emph{entry graph} created by \chemrecon using the example scripts in Sec.~\ref{sec:workflow}.
		The BiGG entry `citrate' is the initial vertex of the graph.
		The light blue vertices represent \te{compound} entries, while the turquoise vertices represent \te{MolStructure} entries, annotated with their confidence scores.
	}
	\label{fig:graph}
\end{figure}

\subsection{Entry Graphs}
\label{sec:entrygraph}
A major advantage of the meta-database structure of \chemrecon is that entries from different sources can be related to each other through relations, as described above.
This creates a massive network of interconnected information. A main feature of \chemrecon is the ability to automatically explore this network and construct subgraphs denoted \emph{entry graphs} of the entries and relations traversed.
These entry graphs can be visualized for intuitive analysis (Fig.~\ref{fig:graph} gives an example), or can be accessed computationally for integration into workflows.

Construction of entry graphs is done by specifying one or more \emph{initial entries} of interest from which the network is then traversed according to a \emph{protocol} defining which entries can be traversed and possibly applying custom Python functions to filter to the discovered entries and relations.
Using these entry graphs, \chemrecon can make connections which are not possible using any single database, or even each database separately.

\chemrecon comes with several pre-defined protocols for a variety of tasks, but also includes a simple way for users to define custom protocols, as demonstrated in Sec.~\ref{sec:workflow}.

As an example, constructing an entry graph starting with the BiGG entry for `isocitrate' and exploring according to a protocol which only allows following the \tr{CompoundReference} and \tr{CompoundHasStructure} relations allow us to explore the full set of molecular structures associated with that compound across all source databases.
The entry graph constructed by this protocol is shown in Fig. \ref{fig:graph}.

\subsubsection{Scoring and Computing Consensus Entries}
\label{sec:consensus}
A key observation behind this project is that the problem of individual database discrepancies can be counteracted by using entry graphs to achieve a high-level overview of the available information associated with an entry across databases.
By leveraging a wide selection of data sources, we can obtain `consensus' information rather than relying on any single source.
An intuitive way to quantify this notion is with `connectedness'---a vertex which is highly connected to the initial entry is more likely to be associated than an isolated vertex, which may be the result of a spurious database connection.

As an example, this idea can be applied on the abovementioned entry graphs of compounds and molecular structures.
Here, the most likely structure (i.e., the consensus structure) for a given compound would be the one most strongly connected to the query compound entry.

We formalize this idea by a \emph{scoring} algorithm which assigns a score to each entry in an entry graph.
The algorithm is based on the PageRank algorithm, and essentially computes the likelihood of a random walk starting at the query entry ending up at each of the entries of the graph.
As with the exploration protocols, the parameters for the scoring algorithm are customizable, for example allowing the user to specify weights for each data source representing the degree to which their information is trusted, which will then be reflected in the final scoring.

We previously implemented the concepts of entry graphs and scoring, restricted to molecular structures, in order to elucidate molecular structures for compounds in metabolic models in \cite{structrecon}.
The entry graph functionality of \chemrecon can be seen as a generalization of this method by vastly expanding the types of biochemical information available.

\subsection{Usage}
\label{sec:workflow}
In this section, we give several examples demonstrating the usage of the Python interface of \chemrecon.
The first feature of \chemrecon is the ability to access database information through a unified interface.
In the following listing, we query \chemrecon for the `isocitrate' entry in the BiGG database, creating a Python object containing the associated information.

\begin{lstlisting}
entry_isocitrate = find_entry(
    id_type = C_BIGG,
    source_id = 'M_icit'
)
\end{lstlisting}

We can use \chemrecon's unified network of connections to obtain more information on an entry.
In this example, we produce a list of \tr{CompoundReference} relations which involve the `citrate' entry.
\begin{lstlisting}
get_relations_from_entry(
    entry = entry_citrate,
    relation_type = CompoundReference
)
\end{lstlisting}

The result is a list of relations, each of which describes the source of the reference and its associated compound entry.

The process of constructing an entry graph applies this procedure iteratively, thereby expanding the set
of related entries.
The following example uses a built-in protocol designed for finding associated molecular structures and applies the scoring algorithm to assign a score to each entry.
Using the \protect\verb|show()| method of the graph then produces Fig.~\ref{fig:graph}.

\begin{lstlisting}
citrate_entrygraph = EntryGraph(
    initial_entries = {entry_citrate}
)
citrate_entrygraph.explore(CompoundStructureProtocol, steps = 4)
scoring = Scorer.score(citrate_entrygraph)
citrate_entrygraph.show(scoring)
\end{lstlisting}

The exploration protocol system is very flexible, and custom protocols can be defined for a wide variety of workflows.
As an example, the listing below defines an exploration protocol which for a given compound explores the set of associated reactions and enzymes in order to find all enzymes which can potentially take the given compound as a substrate.

\begin{lstlisting}
custom_protocol = Protocol(
    relation_types = {
        CompoundReference,
        CompoundParticipatesInReaction,
        ReactionReference,
        ReactionIsCatalyzedByEnzyme,
    },
    relation_types_terminal = {
        EnzymeHasInstance
    }
)
\end{lstlisting}

Creating an entry graph using this protocol results in a graph consisting of the given compound, the compound entries related to it through \tr{CompoundReference}, the reactions in which this compound participates through \tr{CompoundParticipatesInReaction}, the reactions related through \tr{ReactionReference}, and the enzymes which catalyzes the reactions by \tr{ReactionIsCatalyzedByEnzyme}.
Finally, the \tr{EnzymeHasInstance} relations are added to already existing entries (but are not used to explore), revealing hierarchical relationships between the enzymes present.

The above examples only scratch the surface of the functionality.
In particular, most procedures can be customized to support specialized workflows.
The documentation provides further information, including a tutorial, additional examples, and detailed information on the interface.

\section{Discussion}
\label{sec:discussion}
\chemrecon addresses the persistent challenge of integrating heterogenous biochemical data by providing a general, extensible framework, combining a meta-database and unified ontology with a flexible and accessible Python interface.
Although other projects, such as MetaNetX, also consolidate entries from various databases, the approach is different:
MetanetX merges identifiers from various databases and maps them to a single MetaNetX identifier, giving a singular, definitive mapping between these identifiers.
By contrast, \chemrecon makes no such decisions, and simply presents available information and cross-references to the user.
Instead, to associate identifiers, \chemrecon provides the customizable entry graph and scoring functionality, which can be used to discover and visualize \emph{probable} mappings, giving the user full control over the process.

The entry graph abstraction is central to the usefulness of the tool: by representing entries and relations from multiple databases in a single graph, \chemrecon uncovers both direct matches and indirect associations that individual sources cannot reveal, and it enables programmatic exploration of inter-database connections.
Additionally, \chemrecon is a useful tool for easy access to any single database, allowing users to skip the conventional work of implementing HTTP requests or writing parsers.

In summary, we hope that \chemrecon will serve as a useful tool, enabling or streamlining a variety of workflows which rely on database information, and enabling new results using the cross-database analysis and unification functionality of the tool.

%
\section*{Funding}
The work was supported by the Novo Nordisk Foundation grant [NNF21OC0066551].

\section*{Author Contributions}
\textbf{C. A. Eriksen}:
conceptualization,
data curation,
methodology (lead),
software (lead),
writing - original draft.
\textbf{J. L. Andersen}
methodology (supporting),
software (supporting),
supervision,
writing - review \& editing.
\textbf{R. Fagerberg}
methodology (supporting),
supervision,
writing - review \& editing.
\textbf{D. Merkle}
methodology (supporting),
supervision,
writing - review \& editing.

\section*{Availability}
Full documentation, including detailed descriptions of all entries and relations, Python API reference, code examples, and tutorials, is available at \href{chemrecon.org}{chemrecon.org}.
The \chemrecon database is a PostgreSQL 18 database, and is hosted by the Department of Mathematics and Computer Science, University of Southern Denmark.
By default, the \chemrecon Python interface will connect to this database.
Alternatively, the interface can connect to a self-hosted database, available as a Docker image from the homepage.
On most machines, this will be faster, as the latency between the interface and database will be lower.
In this setup, the user can modify the database, for instance by adding custom sources of data.

%
\newpage
\bibliographystyle{plain}
\bibliography{refs}

\end{document}